# Photo-Excitation Dynamics in Electrochemically Charged CdSe Quantum Dots: from Hot Carrier Cooling to Auger Recombination of Negative Trions

*Alireza Honarfar[A]‡, Hassan Mourad[A]‡, Weihua Lin[A], Alexey Polukeev[B], Ahibur Rahaman[A], Mohamed Abdellah[A,C], Pavel Chábera[A], Galina Pankratova[D,E], Lo Gorton[E], Kaibo Zheng[A,F], Tönu Pullerits[A]\**

A. Chemical Physics and NanoLund, Department of Chemistry, Lund University, Box 124, 22100 Lund, Sweden

B. Centre for Analysis and Synthesis, Department of Chemistry, Lund University, Box 124, 22100 Lund, Sweden

C. Department of Chemistry, Qena Faculty of Science, South Valley University, 83523 Qena, Egypt

D. DTU Nanolab, National Centre for Nano Fabrication and Characterization, Technical University of Denmark, Ørsteds Plads, 2800 Kgs. Lyngby

E. Department of Biochemistry and Structural Biology, Lund University, Box 124, 22100 Lund, Sweden

F. Department of Chemistry, Technical University of Denmark, DK-2800 Kongens Lyngby, Denmark

Keywords: Single charged, Double charged, Semiconductor nanocrystals, Spectroelectrochemistry, Transient absorption spectroscopy, Polaron, Tetron, Auger decay, Solar cells, Optoelectronics

**ABSTRACT**: Fulfilling the potential of the colloidal semiconductor quantum dots (QDs) in electrically driven applications remains a challenge largely since operation of such devices involves charged QDs with drastically different photo-physical properties compared to their well-studied neutral counterparts. In this work, the full picture of excited state dynamics in charged CdSe QDs at various time-scales has been revealed via transient absorption spectroscopy combined with electrochemistry as direct manipulation tool to control the negative charging of CdSe QDs. In trions, excited states of single charged QDs, the additional electron in the conduction band speeds up the hot electron cooling by enhanced electron-electron scattering followed by charge redistribution and polaron formation in picoseconds timescale. The trions are finally decayed by Auger process in 500 ps timescale. Double charging in QDs, on the other hand, decelerates the polaron formation process while accelerates the following Auger decay. Our work demonstrates the potential of photo-electrochemistry as a platform for ultrafast spectroscopy of charged species and paves a way for further studies to develop comprehensive knowledge of the photophysical processes in charged QDs more than the well-known Auger decay preparing their use in future optoelectronic applications.

1.INTRODUCTION

Quantum dots (QDs)[1] as semiconductor nanocrystals with size smaller than Bohr radius, have been widely investigated both as model systems for fundamental research and for numerous applications[2–6]. The applications typically rely on separation and extraction of photo-generated electron-hole pairs for the efficient photon to free charge carrier conversion[7]. This can be achieved through transfer of photo-generated charges to electron (hole) acceptors which highly depends on the relative position of the band edges[8]. Charge dynamics in such systems have been extensively investigated by time-resolved spectroscopies[7–9]. Most of these studies are implemented on half-cell systems (i.e. only photoanodes or photocathodes) or open-circuit conditions where electron/hole extraction through external circuits does not occur. In full functioning devices, extraction of charges to/from the electrodes is essential for device operation, which means that under working conditions, the photoactive layer can accumulate charges due to defects or Schottky barrier formation at the interfaces[10]. Charge accumulation at the interfaces and also in the QDs considerably influences the device performance and optoelectronic behavior[11–14]. Therefore, understanding of the effect of an external bias and the presence of extra charges in the QD-acceptor system, is of great importance for the development of high-performance QD devices.

By absorption of light in a semiconductor an electron-hole pair, exciton, is created. Two excitons in a QD interact and form a biexciton, which rapidly decays through Auger process leaving behind a single exciton[15]. Auger recombination in QDs is efficient[16] and drastically influences performance of the corresponding devices at high excitation intensities. Photoexcitation of a charged QD leads to a so called trion which is a three body state of electron-hole pair and an additional charge. Depending on the charge, a trion can be either positive or negative. Analogously to the biexciton, the trion also can decay through Auger recombination[15]. The depopulation of trions through the Auger process competes with the extraction of charges and consequently undermines the charge separation efficiency in the devices. Most of the existing studies on trion dynamics utilized time-resolved photoluminescence spectroscopy to measure radiative trion decay in core/shell heterojunction structures[15,17] where the shell structure has direct effect on the trion recombination[18]. Such core-shell structure isolates the QD core from surface defects, at the same time the shell also slows down or even prevents charge carrier transport to and from the core making integrated optoelectronics based on QDs difficult[19].

In this work we combine electrochemistry with ultrafast transient absorption spectroscopy (TA) to monitor changes in the excited state dynamics of the QD under controlled charging[12,20–23]. By observing the changes in the steady state absorption during spectro-electrochemistry measurement, we have identified distinct potentials which correspond to the injection of one and two electrons to the QDs. The presence of these negative charges leads to negative trions and tetrons upon excitation by laser light in the TA measurement[15]. We observed a series of changes in the relaxation processes due to the extra electrons including nonradiative Auger recombination in 500 ps timescale. We anticipate that our findings will pave the way to comprehensive knowledge and better understanding of the photo-physical processes in charged QDs leading to their efficient use in future nanotechnology applications.

2. ELECTROCHEMISTRY

In electrochemistry the applied potential induces electron exchange between working electrodes and the sample under study[24]. By changing the potential of the working electrode with time (versus a reference electrode) while measuring the current that passes to the counter electrode, a cyclic voltammogram (CV) is obtained[25], showing distinct bands at the potentials where charge carriers can enter the electronic states of the system[14,24,26]. Monodispersed CdSe QDs with a diameter of

~3 nm were used for sensitization of the TiO$_2$ coated FTO (SI). Such TiO$_2$-FTO system is analogous to the photoanode architecture of QD-sensitized solar cells[7]. CVs of CdSe QDs were measured in a conventional three electrode electrochemical cell configuration, where the QD-TiO$_2$-FTO assembly serves as the working electrode, a leakless pseudo Ag│AgCl electrode is used as reference and a platinum wire acts as a counter electrode. Detail description is given in Supporting Information (SI). All the potentials in this work are reported vs. the Ag│AgCl pseudo reference. In equilibrium the Fermi levels of the semiconductors that are in contact with each other are considered to be equal, which will equilibrate the charges through QD-TiO$_2$-FTO layers[27]. In that respect the TiO$_2$-FTO layer can be considered to be as one system with a common Fermi level while CdSe QDs are separate semiconductors attached to the TiO$_2$-FTO by a linker molecule. The reference electrode is considered to have a fixed potential, thereby applying a negative bias corresponds to an increase in the TiO$_2$-FTO Fermi level. All potentials in this paper is reported versus Ag|AgCl pseudo reference (SI). At sufficiently high negative potentials, electrons in the conduction band (CB) of the TiO$_2$-FTO can reach energies equal to or higher than of the CdSe CB, hence electrons can be injected into the QDs. When electrons are exchanged between QDs and TiO$_2$-FTO, a current change can be detected in the CV which is represented in Figure 1A. For a better understanding of electrochemical measurement and changes in the current, it is essential to define the relation between energy levels of the QDs and the applied potentials. By identifying the electrochemical potential for the CB minimum (first excited state) of the QDs $U_{1Se}$ (in Volts), the electrochemical potential for the valence band energy levels U$_h$ $U_h$ can be expressed as[13]

$U_h = U_{1Se} + \Delta E_{opt}/e + 1.8e/(4\pi \varepsilon_r^\infty \varepsilon_0 r)$ ,

where $\Delta E_{opt}$ is the energy of the optical transition from corresponding energy level in the valence band to 1S$_e$, e is the elementary charge, $\varepsilon_r^\infty$ the medium's relative permittivity at high frequencies, $\varepsilon_0$ the vacuum permittivity, and r is the radius of the QDs. The approximate electrochemical potentials for the energy levels obtained from the CV and the discrete energy levels of 3 nm CdSe QDs with their corresponding optical transitions are depicted in Figure 1B and C[13,14,26].

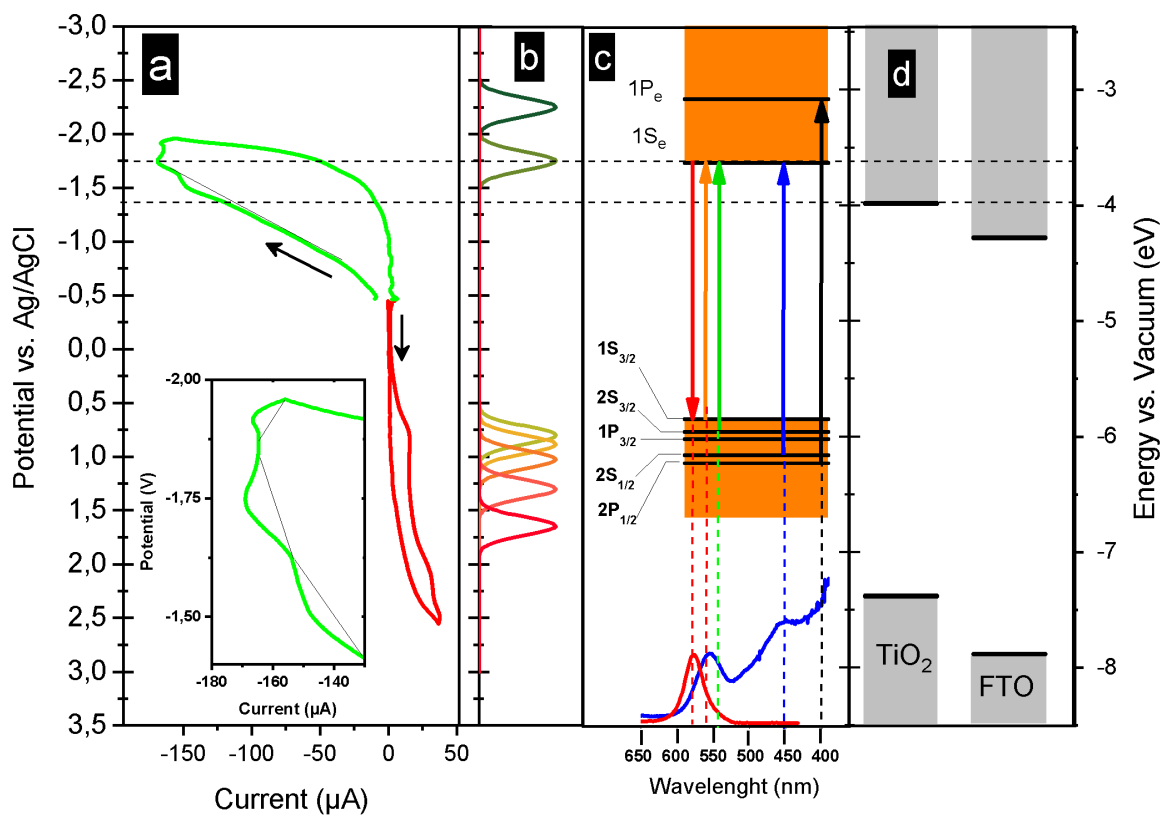

Figure 1: A) Cyclic voltammogram of the 3 nm CdSe QD/TiO$_2$/FTO electrode assembly. Measurement was performed in a three-electrode cell where the CdSe QDs assembly on a TiO$_2$ coated FTO electrode served as the working electrode, Ag/AgCl as the reference and a platinum plate was used as counter electrode. 0.1 M solution of tetra butyl ammonium hexafluorophosphate in dichloromethane was used as the supporting electrolyte. A scan rate of 1 mV/s was used at room temperature. The inset represents a close-up of the current peaks at -1.5 V, 1.75 V and -1.9 V. B)

calculated electrochemical potentials for 3 nm CdSe QDs energy levels by assuming conduction band (first excited state) potential as -1.75 V. C) discrete energy levels of 3 nm CdSe QDs[13,28]. Up and down arrows connecting optically allowed transitions for absorption and emission. In the bottom measured absorption and emission of the 3 nm CdSe QDs are also represented. D) energy levels of the conduction band and valence band of the TiO2 and the FTO.

In CV, Figure 1A, with increasing negative bias we see a steady raise of the current due to Faradaic processes[24]. At -1.5 V, a clear peak emerges since the Fermi level reaches the conduction band of $TiO_2$ and, consequently, additional electrons can be injected into the CB of $TiO_2$. At -1.75 V another peak indicates the electrochemical potential where electrons can be transferred from $TiO_2$ to the $1S_e$ level in the QDs. The next peak at -1.9 V is at too low potential to correspond to the next energy level, $1P_e$, as illustrated in Figure 1B and C [29]. Accordingly, we interpret the peak as further charging of the QDs since the $1S_e$ level can simultaneously accommodate two electrons. Due to the Coulomb repulsion, it is expected that addition of the second electron requires more energy which corresponds to a higher negative potential in electrochemistry. Since the features between - 1.5 V and -1.75 V are a combination of the QDs and that of the TiO2-FTO, exact description of them solely based on CV measurement is not straightforward. In addition, as one can see from CV measurement in Figure 1A and as reported previously in literature, the electrochemical measurements on QDs have irreversible character[14,30,31]. This indicates that the charges injected into the QDs do not fully return on the time scale of the electrochemical cycle when the potential is reversed[30]. Previous studies have shown that charging can induce surface modifications which leads to the formation of traps in QDs without inducing significant changes in absorption and photoluminescence[22]. It has been also shown that it is possible to charge QDs without reducing the surface by using passivating ligand or by building core-shell structures to eliminate such surface species[21,22,32]. For precise scrutiny of the effect of external bias on the QD assembly, as the main goal of this study, combination of electrochemistry with spectroscopy[33] is used. Changes of the absorption spectra due to the state filling can further identify the energy levels that are involved in charging[13,34].

3. SPECTROELECTROCHEMISTRY

The spectroelectrochemistry[35] measurements were carried out by following the absorption spectra at different negative potentials during the CV scan. The initial spectrum was taken at open circuit potential (OCP) corresponding to -0.25 V vs. Ag│AgCl and the potential dependent spectral changes are obtained by subtracting this initial spectrum from the measured absorption spectra at each potential as represented in Figure 2.

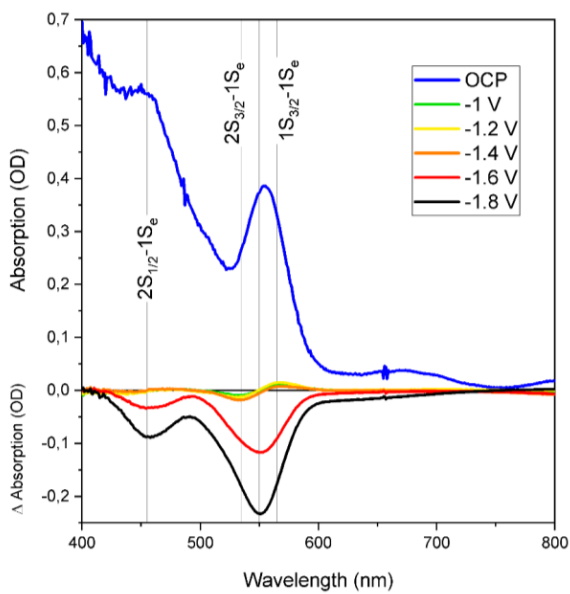

Figure 2: Spectroelectrochemical steady state absorption changes of 3 nm CdSe QDs in $TiO_2$/FTO. Upper part: steady state absorption of 3 nm CdSe QDs in $TiO_2$/FTO where no potential is applied (OCP = -0.25 V vs. Ag/AgCl). Lower part: potential dependent difference spectra obtained by subtraction of the OCP spectrum from the measured spectra. All potentials are measured versus Ag|AgCl pseudo reference, and a platinum plate was used as counter electrode. 0.1 M solution of tetra butyl ammonium hexafluorophosphate in dichloromethane was used as the supporting electrolyte.

The absorption spectra show two clear bands at around 550 nm for $1S_{3/2}$-$1S_e$ and $2S_{3/2}$-$1S_e$ transitions and at 450 nm for the $1S_{1/2}$-$1S_e$ transition. At all negative potentials, an increasing featureless background from 600 nm to 1000 nm (Figure S5) is due to absorption of the charges in $TiO_2$ [36]. Until the potential reaches -1.5 V, the Fermi level in the $TiO_2$-FTO layer is lower than the QD conduction band, hence the induced absorption changes at 535 nm and 565 nm are due to the Stark shift[37]. At -1.6 V and higher negative bias, strong bleach of the transitions at 550 nm and 450 nm show that the electrons fill the QD $1S_e$ state and consequently the absorption of all transitions to this level become partially bleached. At -1.8 V, the bleach amplitude becomes two times larger. These observations provide strong evidence that at -1.6 V the QDs are charged by one, and at -1.8 V by two electrons which fills the doubly degenerate $1S_e$ state. These results demonstrate that spectroelectrochemistry provides a convenient direct manipulation tool for a controlled charge-injection to the QDs. In the following we will be using this method to prepare negatively charged QDs for the transient absorption measurements. We also point out that we are well aware of the limitations of the effective mass theory description and in reality the colloidal QDs have much more energy levels shown by atomistic calculations[38]. Such a quasi-continuum view is important for understanding the excited electron relaxations in QDs in the following. However, the main spectral features of these two approaches are in good agreement justifying our interpretation of electrochemistry in terms of k•p effective mass theory and state filling. It is worth mentioning that even at -1.8 V the band edge transition is not fully bleached. This may be due to a subset of QDs with incomplete charging. It may also be the result of the size distribution of the QDs as the smaller dots are less likely to become fully charged. This means that in reality at -1.8V there is a mixture of states which is making precise distinction of the processes in the following analyses complicated and the proposed models need to be taken as simplification of the full complexity of the real situation.

4. TRANSIENT ABSORPTION SPECTROSCOPY

Combination of spectroelectrochemistry with time resolved femtosecond transient absorption spectroscopy (TA) allows *in situ* probing of the excited state dynamics in well-controlled charged QDs. A detailed description of the TA measurements is given in SI. Excitation at 400 nm was used to populate higher energy levels ($2P_{1/2}$-$1P_e$ transition) so that the absorption changes under different electrochemical potentials do not affect the excitation conditions. The initial hot electron relaxation provides valuable information about the details of the carrier cooling process. Decay associated spectra (DAS) were obtained from global fitting of the TA data and used for mapping of the population and depopulation pathways of the excited states. The DAS spectra without normalization are presented in the SI. The TA measurements at OCP (-0.25V vs. Ag/AgCl) are taken as reference. In the following the TA features are discussed in consecutive order.

4.1. OCP AND NEGATIVE BIAS

Relaxation of the initially excited hot electrons and holes occurs in sub-ps timescale[39]. All DAS components of TA consist of a negative peak at 550nm corresponding to a decay of the bleach signal related to the almost instantaneous state filling of $1S_{3/2}$-$1S_e$ and $2S_{3/2}$-$1S_e$ transitions. Since at OCP, after the laser excitation only one electron and hole are present in QDs, electron-electron scattering cannot take place hence the relaxation is induced only by electron-phonon scattering. Since the phonon frequencies are significantly smaller than the level spacing in the effective mass description, the hot carrier cooling corresponds to relaxation through a quasi-continuum of the energy levels[38]. The small positive band at 600 nm in the 500 fs DAS is excitation-induced spectral shift and also reflects the hot carrier arrival to the band edge[40,41]. It follows with two slower processes corresponding to the initial electron injection (9 ps) to a charge transfer state where the electron still interacts with the hole and the full charge separation via electron diffusion to the bulk of the $TiO_2$ (95 ps)[42]. The broad positive signal at longer wavelengths is attributed to the excited state absorption (ESA) of free electrons in the CB of $TiO_2$ [36]. Eventually, the photo-generated charges are being recovered within a timescale that is longer than our experimental limit noted as >10 ns in all cartoon illustrations of the decay processes in Figure 3.

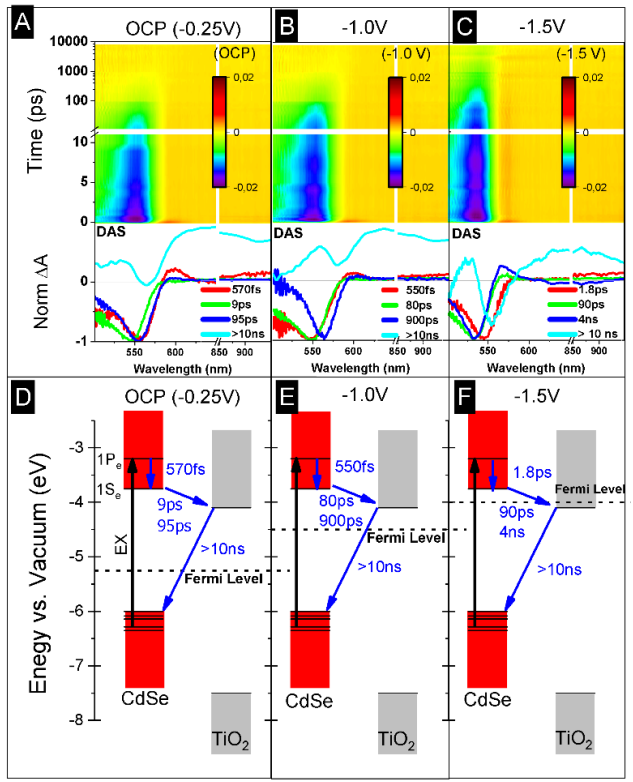

Figure 3: A, B and C: Summary of TA measurements and their corresponding fitted exponents with normalized decay associated spectra DAS of the 3 nm CdSe QD assembly on TiO$_2$ at OCP, -1 V and -1.5 V applied bias versus Ag|AgCl, respectively. The corresponding unnormalized DAS component is shown in SI. D, E and F: Cartoon depiction of electron transitions at the given potentials. Black arrows indicate laser excitation connecting corresponding energy levels for photo-generated charges, blue down arrows are used to show electron decaying pathways and recombination mechanisms.

### 4.2. SINGLE AND DOUBLE CHARGED QDs

When an external voltage is being applied to the photoanode, at potentials up to -1.5 V no electrons are electrochemically injected into the QDs, but the excess electrons in the TiO$_2$-FTO do affect the excited state dynamics in the QDs as well as charge transfer to TiO$_2$ due to long range Coulomb interaction in QDs[43,44]. At -1.0 V the Fermi level is still below the CB of the TiO$_2$ and the electrons can only enter the TiO$_2$ shallow traps[45]. The timescale of the hot electron relaxation is not influenced by this, while the photo-excited electron injection from the QDs to the TiO$_2$ and the following diffusion of the injected electron away from the QD vicinity into the bulk of TiO$_2$ are both significantly slowed down from 9ps to 80 ps and from 95ps to 900 ps, respectively. At -1.5 V, when electrons fill the lowest levels of the TiO$_2$ conduction band, the charge separation process is further slowed down up to ns. The fast component that was earlier related to the hot carrier cooling, has also become significantly slower representing a combination of the cooling and some other slower process.

When the applied negative bias reaches -1.75 V, the Fermi level of the TiO$_2$-FTO is shifted higher than the QD 1S$_e$ level which becomes populated by one electron. Excitation of such charged QDs creates trions. The hot electron relaxation in the trion is shortened to 370 fs due to the electron-electron scattering, which is now possible because of the extra electron[41]. Since at - 1.75 V, the electron density in the CB of the TiO$_2$ is high, the electron injection from QDs to TiO$_2$ is very unlikely to occur. The 5.8 ps component is interpreted as a combined effect of Coulombic repulsion between the electrons and polaron formation[40]. The repulsion is likely to push the extra electron to the surface of the QDs. Such a new charge distribution is followed by nuclear rearrangement forming a polaron and thereby stabilizing the charge distribution[46]. Observation of strong Stokes shift of the fluorescence in electrochemically single and double charged CdSe QDs gives a strong evidence for polaron formation and for the related nuclear rearrangement[20]. The following 510 ps component is assigned to negative trion Auger recombination (see Figure 4 A and C) which is in good agreement with the previous reports[15,18].

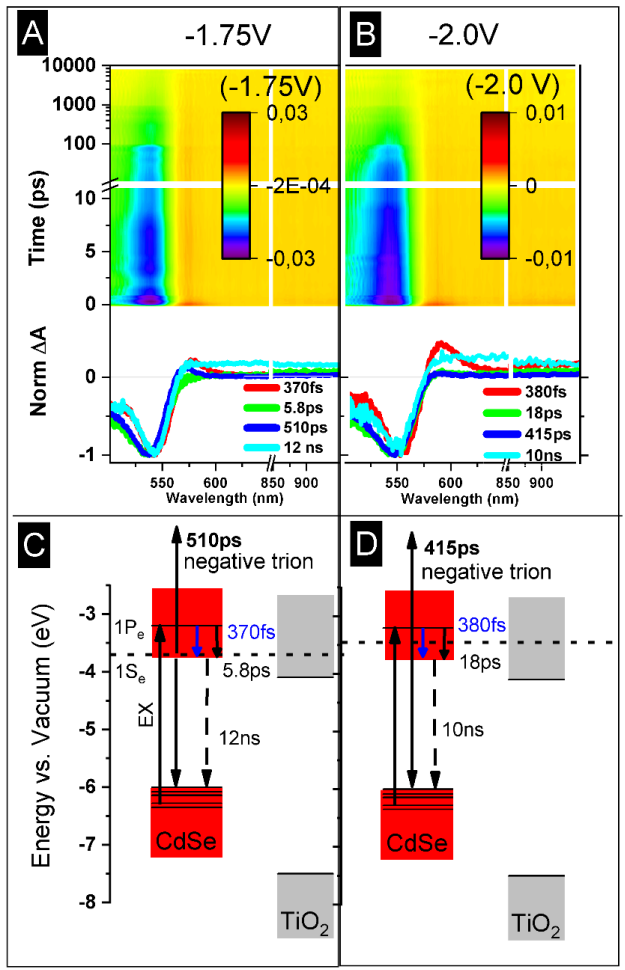

Figure 4: A and B. summary of TA measurement and their corresponding fitted exponents with decay associated spectra for -1.75 V and -2.0 V, respectively. The potentials are in respect to the Ag/AgCl reference electrode. C and D. Cartoon depiction of electron transitions, where black up arrows indicate laser excitation connecting corresponding energy levels for photo-generated charges, blue down arrows are used to show electron decaying pathways and recombination mechanisms. The Auger process is illustrated by double head arrows to reflect the simultaneous recombination of electron-hole and excitation of an electron.

After the Auger recombination, the QDs return back to the initial unexcited charged state, but the DAS analysis still shows a 12 ns component (10% of the TA signal SI Table S2) with a clear ground state bleach at the main absorption band. This minor long component corresponds to the QDs which have remained uncharged.

At -2 V, since QDs become double charged, the corresponding excited states are called tetrons, three electrons and a hole[47]. The hot electron relaxation (380fs) has about the same timescale as in case of the negative trions. The relaxation due to polaron formation becomes longer since stronger electrostatic repulsion can lead to larger charge redistribution and the corresponding rearrangement of the QDs' lattice needs more time to reach the new minimum energy level. As expected, the Auger process in the tetron is slightly shorter (415ps) than in trion.

## 5. CONCLUSION

In conclusion, spectroelectrochemistry is used to directly manipulate and monitor charging of the QD photoanode assembly in a well-controlled manner. Combination of spectroelectrochemistry with *in situ* TA provided a convenient tool for study of the excited state dynamics of the QDs under different potentials mimicking realistic working conditions in optoelectronic device. At low potentials the electrons cannot reach the CB of the $TiO_2$ but populate the shallow traps which appear as constant featureless background over the whole visible spectral range. Next, the electrons start filling the CB of $TiO_2$ without entering the QDs and the photo-induced electron transfer from the QDs to TiO2 is significantly slowed down. At even higher negative potentials, -1.75 V and -2.0 V, spectroelectrochemistry confirmed that the QDs are charged by one and two electrons, respectively. Now the hot electron cooling process becomes faster due to the electron-electron scattering. In the charged QDs a process in the picosecond timescale occurs, which is attributed to polaron formation. In addition, Auger relaxation with 500 ps time constant is observed in our TA measurements. The Auger process becomes faster in case of double charging, in tetrons. Thereby, we provide a thorough *in situ* spectroscopic investigation of QDs at operational conditions of devices which depend on charge separation or electro-charging of the particles.

The work can lead to a novel photophysical view on the design of QDs in optoelectronic devices. Our exploration of the effect of slight variation of charging that cause diminishing of charge extraction provide understanding of the photophysical ground for efficiency limits of the QD devices.

## ASSOCIATED CONTENT

**Supporting Information** including Materials and chemicals, QDS synthesis, Film preparation, Electrochemical measurement, Spectro-electrochemistry, Transient absorption spectroscopy are available in the following files free of charge.

supporting information.PDF

## AUTHOR INFORMATION

**Corresponding Author**

*Tönu Pullerits tonu.pullerits@chemphys.lu.se   ORCID: 0000-0003-1428-5564

Chemical Physics and NanoLund, Department of Chemistry, Lund University, Box 124, 22100 Lund, Sweden

Visiting & Deliveries Address: Kemicentrum, Naturvetarvägen 16, 22362 Lund, Sweden

Postal Address: Chemical Physics, P.O. Box 124, SE-22100 Lund, Sweden

Phone +46 46 22281310    Fax  +46 46 2224119**Present Addresses**

†If an author's address is different than the one given in the affiliation line, this information may be included here.

**Author Contributions**

The manuscript was written through contributions of all authors. All authors have given approval to the final version of the manuscript. ‡These authors contributed equally.

**Funding Sources**

K. Z. acknowledges financial support from the Independent Research Fund Denmark-Sapere Aude starting grant (No. 7026-00037A) and Swedish Research Council VR starting grant (No. 2017-05337). W. L. thanks the financial support by China Scholarship Council (CSC) under grant No. 201806460021.

ACKNOWLEDGMENT

Authors acknowledge funding by the Swedish Research Council, Swedish Energy Agency, KAW foundation and NanoLund. A. H. thanks Prof. Ebbe Nordlander, Prof. Ola F. Wendt and Dr. Carlito S. Ponseca for their scientific support.ABBREVIATIONS

QD, Quantum dot; CB, conduction band; CV, cyclic voltammogram; 3MPA , 3-mercaptopropionic acid; FTO, fluorine doped tin oxide;  ITO, indium doped tin oxide;  HOMO, highest occupied molecular orbital; LUMO, lowest unoccupied molecular orbital; OCP, open circuit potential; TA, transient absorption spectroscopy; DAS, Decay associated spectra. ESA, excited state absorption.

Insert Table of Contents artwork here

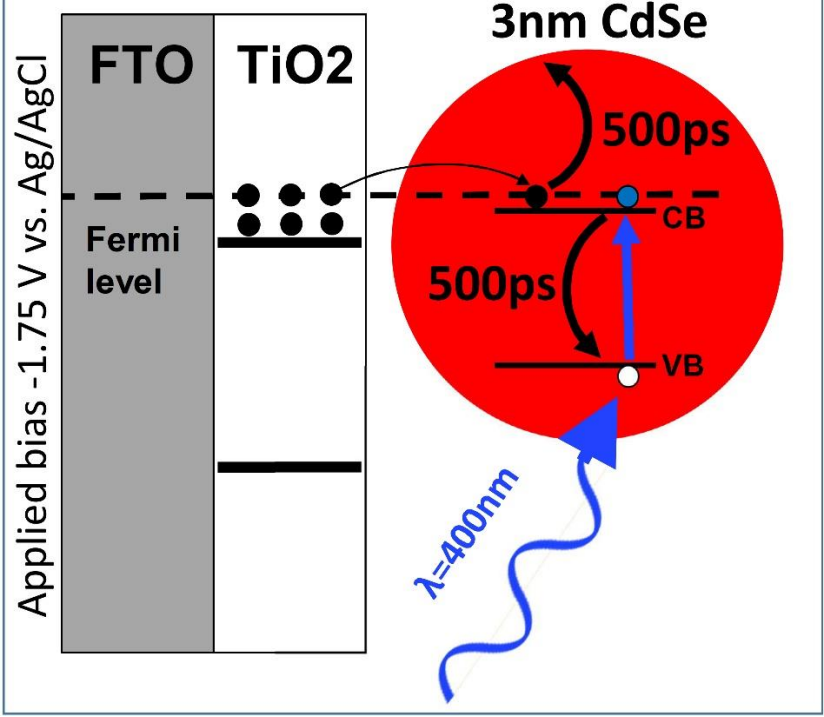

**Supporting information:**

**Photo-Excitation Dynamics in Electrochemically Charged CdSe Quantum Dots: from Hot Carrier Cooling to Auger Recombination of Negative Trions**

Alireza Honarfar[A,‡], Hassan Mourad[A], Weihua Lin[A], Alexey Polukeev[B], Ahibur Rahaman[A], Mohamed Abdellah[A,C], Pavel Chábera[A], Galina Pankratova[D,E], Lo Gorton[E], Kaibo Zheng[A,F], Tönu Pullerits*[A]

- A  Chemical Physics and NanoLund, Department of Chemistry, Lund University, Box 124, 22100 Lund, Sweden

- B  Centre for Analysis and Synthesis, Department of Chemistry, Lund University, Box 124, 22100 Lund, Sweden

- C  Department of Chemistry, Qena Faculty of Science, South Valley University, 83523 Qena, Egypt

- D  DTU Nanolab, National Centre for Nano Fabrication and Characterization, Technical University of Denmark, Ørsteds Plads, 2800 Kgs. Lyngby

- E  Department of Biochemistry and Structural Biology, Lund University, Box 124, 22100 Lund, Sweden

- F  Department of Chemistry, Technical University of Denmark, DK-2800 Kongens Lyngby, Denmark

**Corresponding Author**

*Tönu Pullerits tonu.pullerits@chemphys.lu.se   ORCID: 0000-0003-1428-5564

Chemical Physics and NanoLund, Department of Chemistry, Lund University, Box 124, 22100 Lund, Sweden

Contents

Page

I. Materials and chemicals

Octadecene (ODE), Cadmium oxide (CdO), Selenium powder (Se), Oleic acid (OLEA), Trioctylphosphine (TOP), Dichloromethane (DCM), Methanol (MEOH), Acetone, 3-mercaptopropionic acid (3-MPA), Fluorine doped tin oxide on glass (FTO), tetrabutylammonium hexafluorophosphate (TBAPF6). All chemicals and materials were purchased from Sigma-Aldrich and were used as received without any purification or distillation. 3Å molecular sieves powder was activated by heating under vacuum then used for drying and storage of solvents, if needed.

II. QD synthesis

For the synthesis of CdSe nanoparticles, a conventional hot injection method was used with slight modification [1]. For a typical synthesis of CdSe nanoparticles, 1500 mg of CdO was dissolved in 7 ml of OLEA and 50 ml of ODE at 270°C in a three-neck flask. Selenium precursor was prepared by sonication of Se powder in 10 ml of ODE for 10 min in an Ar purged flask then 0.5 mL of Trioctylphosphine was added and then stirred until it became a transparent solution. When the Cd precursor solution became clearly transparent the temperature was lowered to 240°C, then the Se solution was quickly injected at 240°c and stirred for 2 min. Then the flask was removed from the heater and the hot solution was quickly poured into a metallic bucket, which was placed in a cold bath filled with dry ice. This immediate cooling proved to be very efficient to keep the size distribution of the nanoparticles very narrow. For purification a modified method was used as given elsewhere [2], dichloromethane was used as solvent instead of toluene. The reason for using DCM is that it has a higher density, so it is a very good solvent for extraction of CdSe nanoparticles from the mixture of Cd-oleate in methanol and acetone.

Since the presence of excess oleic acid can complicate the ligand exchange, $^1$H NMR was used to confirm the success of the washing process (see Fig SI 1.). The synthesized CdSe QDs were characterized to ensure the quality, size distribution and shape of the nanoparticles. The absorption spectrum shows a very distinctive excitonic peak around 555 nm. By calculating the concentration of the QDs in solution and measuring the absorbance at the excitonic peak, we evaluated the size of the QDs to be 3.5 nm. The SEM image shows spherical nanocrystals with a relatively narrow size distribution, see Fig. SI 1. Photoluminescence of the QDs excited at 400 nm, shows a narrow emission band at 575 nm with FWHM 30 nm, which is comparable with the reported literature data [3]. The size distribution of the QDs has a significant effect on the electrochemistry and charge transfer [4] behavior of the system under study.

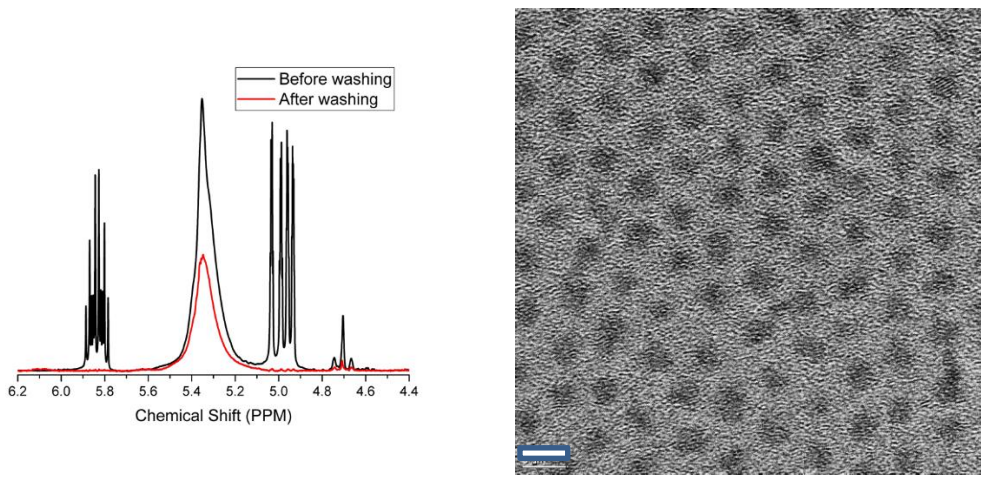

*Figure S1. $^1$H NMR of washed and unwashed QDs, TEM image of the QDS scale bar 5nm*

**III.** Film preparation

By a conventional ligand exchange method, the capping ligand was changed to 3-MPA. The FTO slides were coated with TiO$_2$ paste, sintered at 480°C and cooled overnight[5]. One should consider that annealing of the TiO$_2$ can have significant effect on electrode properties. TiO$_2$ mesoporous layer was to achieve large surface area for QDs anchoring and thereby obtaining sufficient optical density for measurements. Films were prepared by soaking TiO$_2$ coated FTO slides in a MPA capped QDs solution at basic pH>10 for more than 24 h, then washed with distilled water and heated to 110°C. Films were stored in a vacuum desiccator filled with Ar over drying agents to make sure it will be dry by the time of the experiments. Films were kept in the dark, to avoid light soaking and photodegradation was observed for samples under illumination due to charge transfer [6].

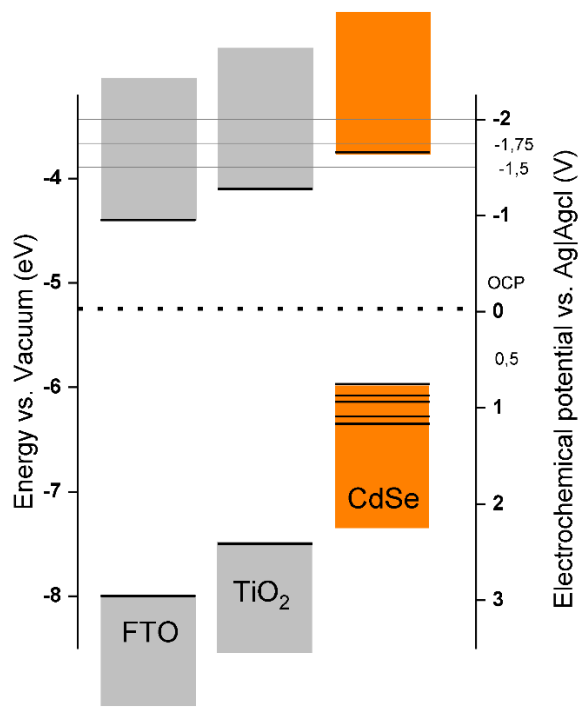

*Figure S2. the schematic band alignment of 3nm CdSe QDs on TiO2 in electrochemical cell alignments*

## IV. Electrochemistry measurement

A conventional 3 electrode system was used. TiO$_2$ coated FTO glass was used as working electrode, Pt wire as counter electrode and leak less Ag☐AgCl electrode as reference electrode [7]. The electrochemical cell was a homemade cell made from glass with a 3 mm spacer. For electrolyte solution 0.1M tetrabutylammonium hexafluorophosphate in DCM was used. The direction of the CV scans is indicated with arrow in the plotted CV. A scan rate of 1 mV/s was employed to ensure quasi-stationary conditions. The corresponding stationary open circuit potential (OCP) is -0.25 V vs. Ag/AgCl. Palmsens 4 potentiostat was used for electrochemical measurements. Electrolyte selection is an extremely critical decision, because adsorption and intercalation of electrolyte molecules into the nanoparticles shows significant effects on the electrochemistry [8]. To perform reliable electrochemical measurement on QDs, one should pay attention to the possible degradation reactions. One of the most important challenges is the effect of oxygen and water. In the course of our experiments we realized that LiClO$_4$ as supporting electrolyte containing perchlorate are the worse option to be used, because upon decomposition it provides very large quantities of oxygen inside the cell. Also, using some chemicals like DMF is not advisable [9]. After thorough considerations and preliminary experiments, we choose DCM as solvent and 0.1M TBAHFP for supporting electrolyte. DCM is not hygroscopic, which reduces the presence of water in our experiments significantly and also it is a good solvent of organic compounds. In below control measurements of CV of blank FTO and TiO$_2$ coated FTO are presented. In the scan toward more negative potentials, more current corresponds to further charging of the TiO$_2$-FTO since the density of states of the TiO$_2$ conduction band is significant and the raise of the current shows that more electrons can enter into the TiO$_2$ layer at a higher potential (Figure SI 2).

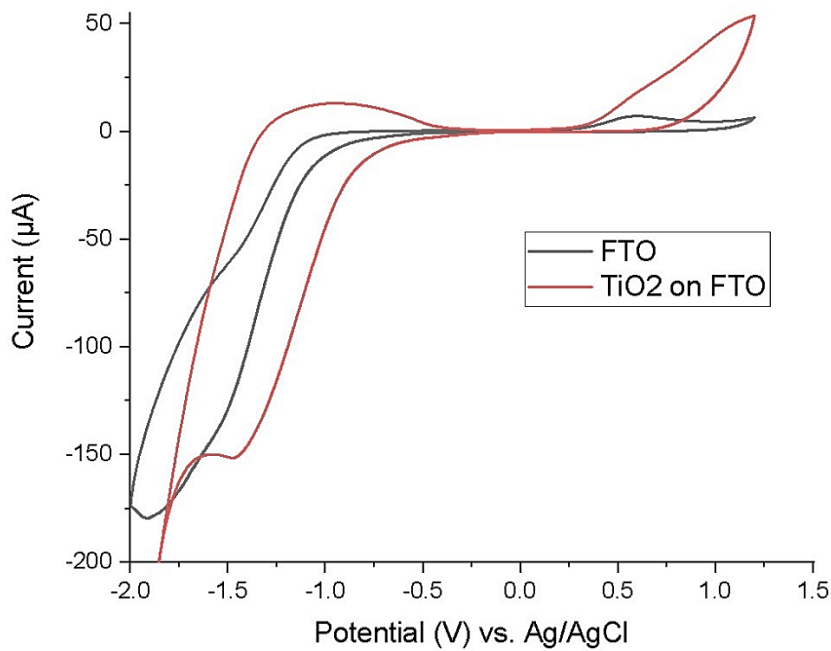

*Figure S3. CV of blank FTO and TiO2 coated FTO*

V.   Spectro-electrochemistry

A very slow scan rate of 1 mV/s was used to record UV-vis absorbance spectra under quasi stationery conditions. Spectral measurements were done at a limited number of specific potentials to prevent photo charging or photo currents to cause peaks in CV, which can further complicate the interpretation. The acquisition of a spectrum took about 5 second causing a negligible effect on the current and during which the potential changes only by ±5 mV. Spectral measurements were done at limited number of specific potentials to prevent photo charging or photo currents to overlap with the peaks in CV, which can further complicate the interpretation. In steady state measurement, presence of additional electrons causes changes in the absorption spectrum. For example, excitonic peaks bleach and absorption of higher energy levels can appear. Similar changes occur in TA measurement. The results are presented in Fig. SI 5.

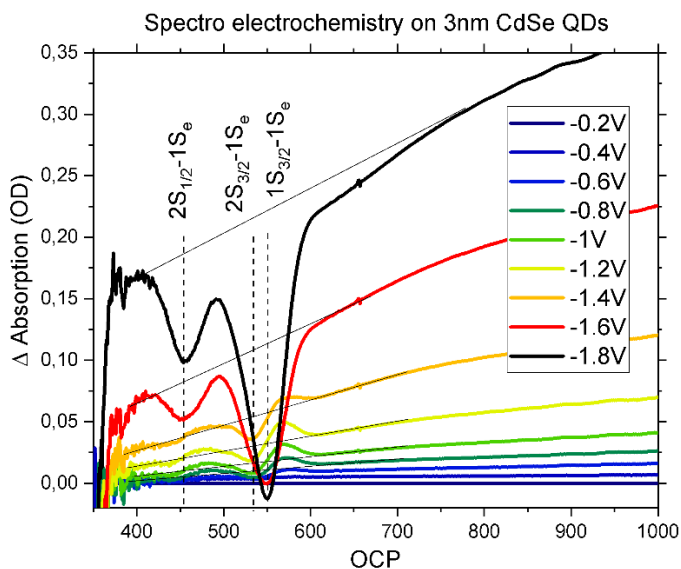

*Figure S4. Change in steady state absorption of QDs under application of negative bias. The absorption due to the electrons in $TiO_2$ is represented by lines and was substracted to make the Fig 2 of the main text.*

**VI.** Transient absorption spectroscopy

Output pulses of regenerative amplifier Spitfire Pro, 796 nm, 6mJ, 100fs were used to generate both pump and probe light. Second harmonic of the fundamental laser wavelength generated by BBO crystal was used as pump pulses with central wavelength 400 nm to ensure that the photo-excited electrons reach higher energies of the conduction band and are not disturbed by the electrochemically injected electrons. Output of a NOPA (Topas) was used to generate 1300nm which by CaF$_2$ crystal generate broadband white light as probe covering the range from 350 nm to 1200 nm. After setting a specific potential and reaching equilibrium, spectral evolutions of the photoexcited QDs were recorded to track the excited state dynamics.

The data are collected as average of 1000 laser pulses at each delay point. In our measurement the negative signal corresponds to the ground state bleach and stimulated emission while the positive signal is excited state absorption. In order to avoid the multiexciton effects on the dynamics, a very low photon flux was used (50μW at 400μm spot size). For Transient absorption spectroscopy we used bulk-electrolysis or chronoamperometric method. In this method a potential in set for the whole period of the measurement. Only after the current is stabilized, we performed laser measurements. For example, in Fig. SI 4., we can see that after 100 second the current does not have any rapid changes and laser measurement can be started. Earlier than that the conditions would not be stable enough for the measurements because the charging of the redox species is not equilibrated.

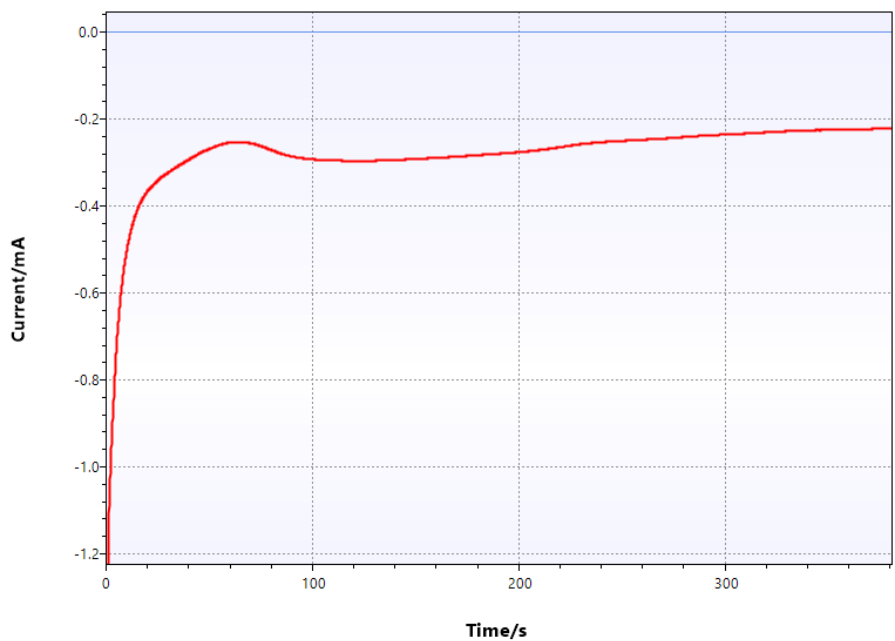

*Figure S5. Chronoamperometric measurement of the QDs sensitized film.*

Global fitting was used as main data analyses method to model the time dependence of the experimental data as a sum of exponential decays by using Glotaran software. Instrument response function was considered to be a Gaussian with FWHM 60fs in the fitting model. In the following we present the decay associated spectra for each data set as obtained from fitting. The onset of the bleach occurs on the timescale of our apparatus response function and we will not be further interpreted and considered.

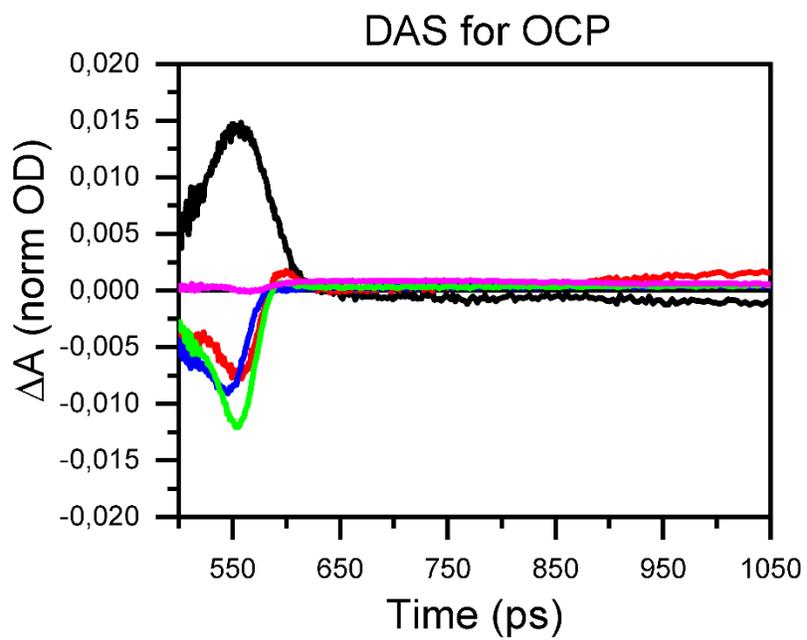

*Figure S6. OCP decay associated black=163.473 fs ,red=570.701 fs ,blue=9.17633ps ,green=95.5214ps ,purple=534.607 ns*

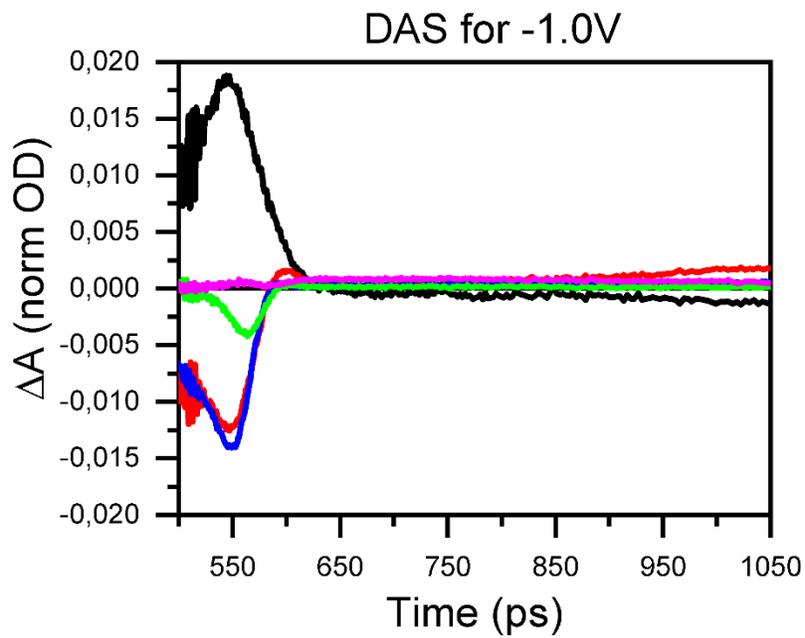

*Figure S7. -1V black=170.899 fs, red=558.569 fs, blue=79.8805ps, green=900.907ps, purple=552.423 ns*

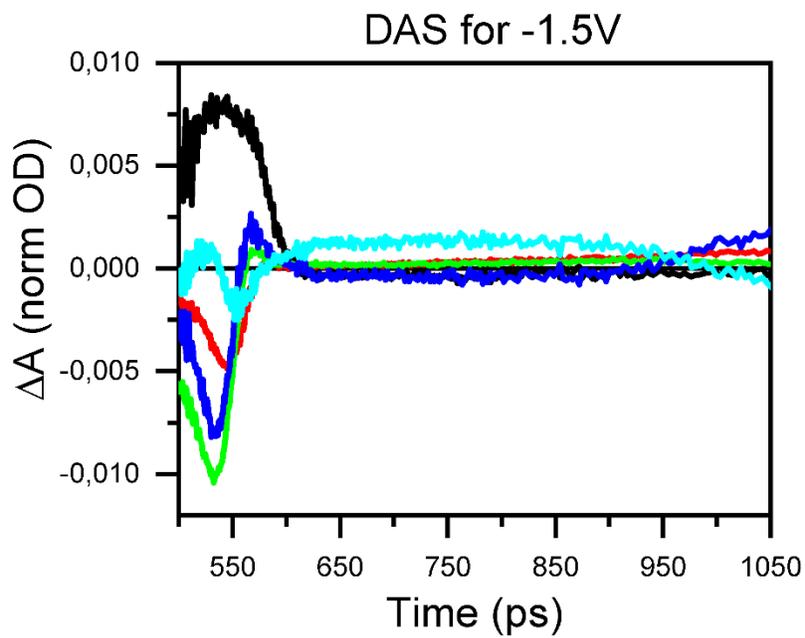

*Figure S8. -1.5V black=147.867 fs, red=1.87586ps, blue=90.3628ps, green=4.03509 ns, cyan=10.8982 ns*

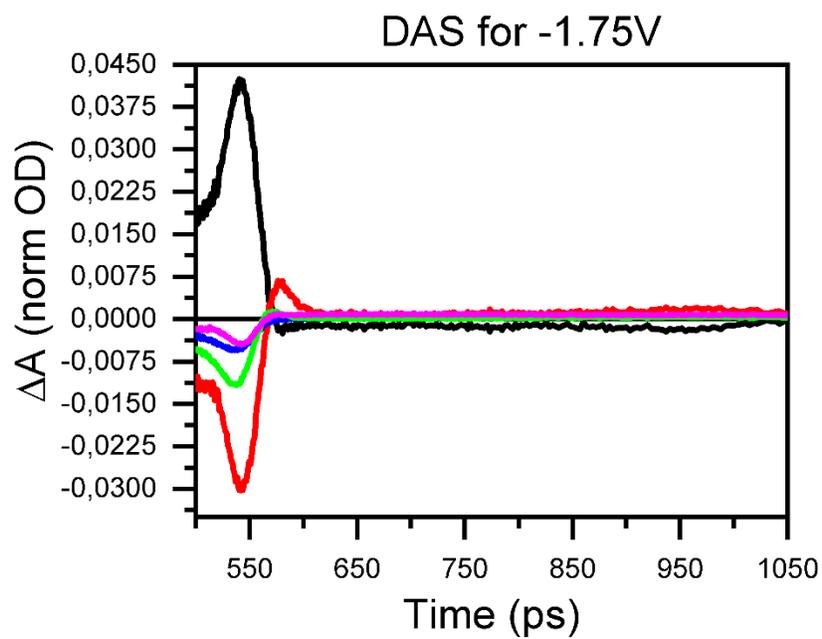

*Figure S9. -1.75V black=177.541 fs, red=369.092 fs, blue=5.81781 ps, green=510.41ps, purple=12.0625 ns*

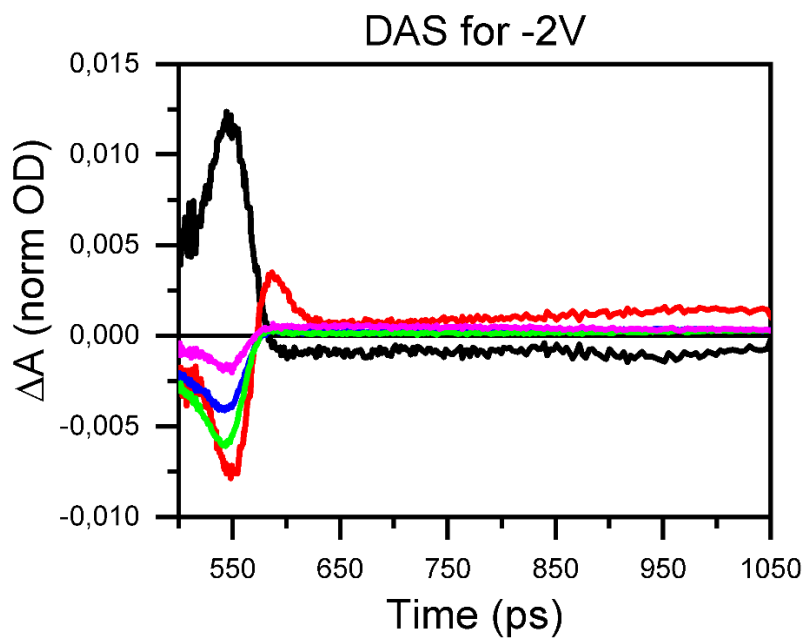

*Figure S10. -2V black=148.417 fs, red=384.726 fs, blue=18.4396 ps, green=415.348 ps, purple=10.3995 ns*

Figures SI 11 to SI 15 show the kinetic traces and the fits at specific probe wavelengths without normalization. For better visualization, the experimental data are represented with line and symbols and only each 5th point is shown. The fit for each trace is plotted as solid line with the same color.

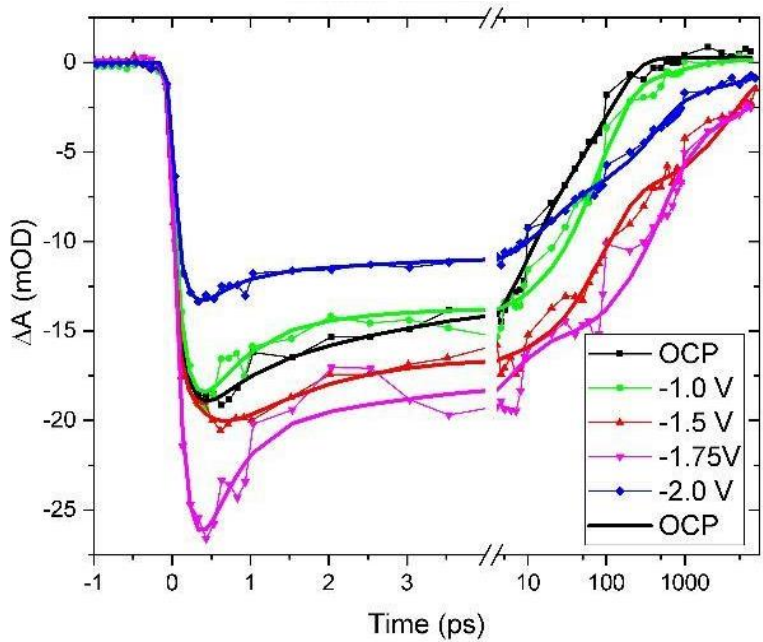

*Figure S11. 540nm kinetic traces for negative potentials as dotted plots with symbols with their corresponding fit as solid line*

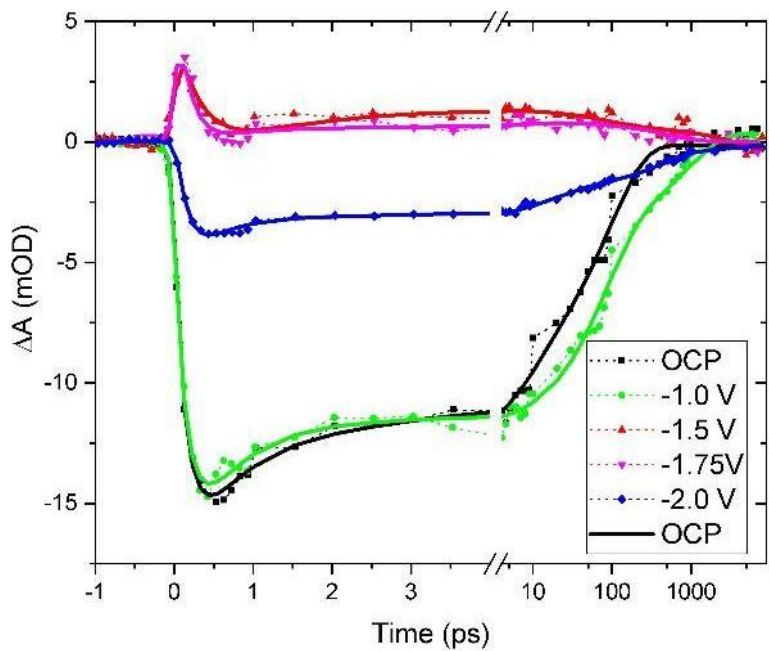

*Figure S12. 566nm kinetic traces for negative potentials as dotted plots with symbols with their corresponding fit as solid line*

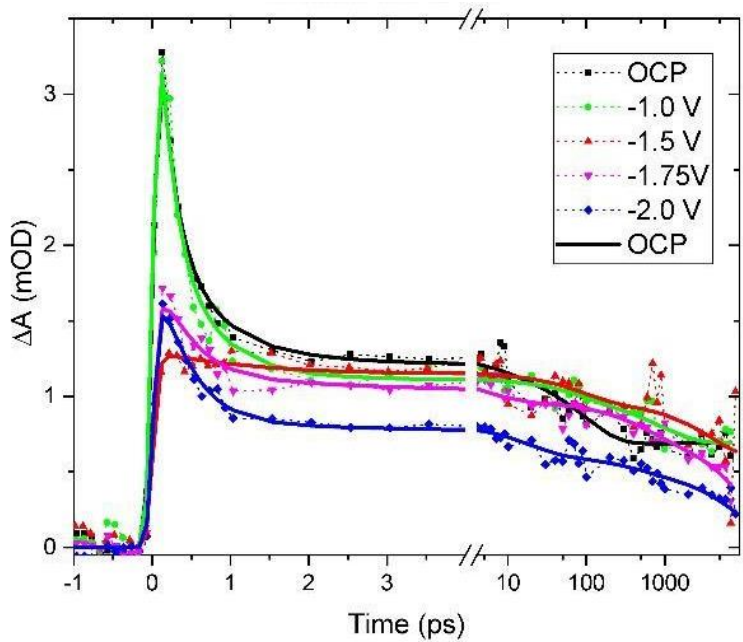

*Figure S13. 607nm kinetic traces for negative potentials as dotted plots with symbols with their corresponding fit as solid line*

The onset of the bleach occurs on the timescale of our apparatus response function and we will not be further interpreting and considering dynamics faster than 170fs. The charge carriers are mainly originate from the direct excitation of the CdSe by the 400 nm laser, therefore the instant onset of the signal within the apparatus response function at the red part of the spectrum(for example 850nm given in below plot). Since no clear rising component of this relatively week signal can be extracted, we conclude that the electron injection to TiO2 is hot electron transfer [10].

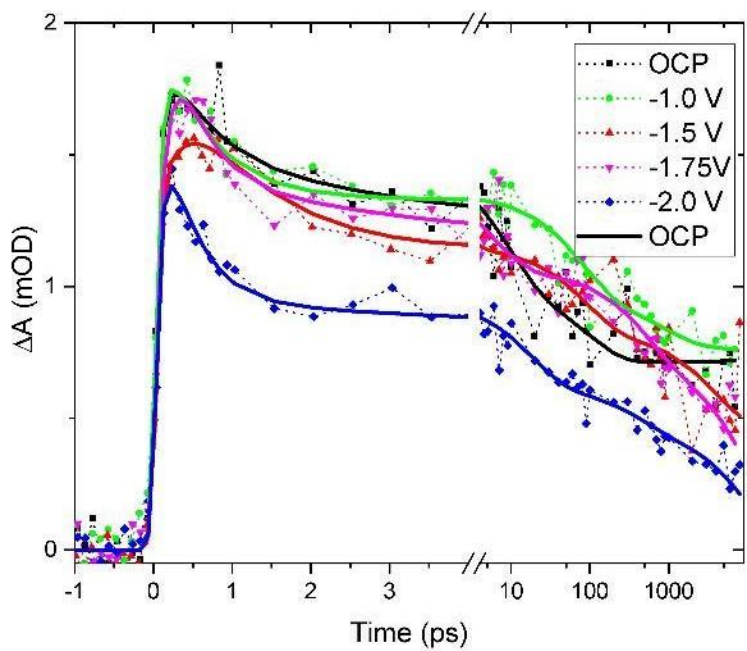

*Figure S14. 850nm kinetic traces for negative potentials as dotted plots with symbols with their corresponding fit as solid line*

*Table S1: Summary of decay lifetimes with their corresponding contribution to the negative signal in TA measurements*

| Plot | t1 | t2 | t3 | t4 |
|------|------|------|------|------|
| OCP  | 570 fs (27%) | 9 ps (31%) | 95 ps (42%) | >10 ns (0.035%) |
| -1.0 V | 550 fs (41%) | 80 ps (46%) | 900 ps (13%) | >10 ns (0.033%) |
| -1.5 V | 1.8 ps (18%) | 90 ps (42%) | 4 ns (32%) | >10 ns (8%) |

*Table S2: Summary of the decay components with their corresponding contribution to the negative signal in TA measurements*

| Plot | t1 | t2 | t3 | t4 |
|---|---|---|---|---|
| -1.75 V | 370 fs (58%) | 5.8 ps (11%) | 510 ps (23%) | 12 ns (8%) |
| -2.0 V | 380 fs (39%) | 18 ps (21%) | 415 ps (31%) | 10 ns (9%) |